\documentclass[runningheads]{llncs}
\usepackage{graphicx}

\usepackage[ruled,vlined]{algorithm2e}
\usepackage{algorithmic}

\usepackage{color,soul}

\newcommand{\martin}[1]{\textcolor{blue}{\textbf{Martin: }#1}}

\begin{document}
\title{Data Centred Intelligent Geosciences: \\ Research Agenda and Opportunities \\ Position Paper\thanks{This work is funded by the project ADAGEO, IEA CNRS collaboration with Federal University of Rio Grande do Norte  \url{https://adageo.github.io}.}}
\titlerunning{Data Centred Intelligent Geosciences}
%
\author{
Aderson Farias do Nascimento\inst{1} 
\and
Martin A. Musicante\inst{1} 
\and
Umberto Souza da Costa\inst{1} 
\and
Bruno M. Carvalho\inst{1} \and
Marcus Alexandre Nunes\inst{1} \and
Genoveva Vargas-Solar\inst{2} 
}
\authorrunning{A. Farias do Nascimento et al.}
%
\institute{
Universidade Rio Grande do Norte, Brazil \\
\email{aderson@geofisica.ufrn.br,mam@dimap.ufrn.br,umberto@dimap.ufrn.br, bruno@dimap.ufrn.br, marcus.nunes@gmail.com} 
 \and
 CNRS, Univ Lyon, INSA Lyon, UCBL, LIRIS, UMR5205,69622 Villeurbanne, France \\
 \email{genoveva.vargas-solar@cnrs.fr} 
}

\maketitle              
\begin{abstract}
This paper describes and discusses our vision to develop and reason about best practices and novel ways of curating data-centric geosciences knowledge (data, experiments, models, methods, conclusions, and interpretations). This knowledge is produced from applying statistical modelling, Machine Learning, and modern data analytics methods on geo-data collections. The problems address open methodological questions in model building, models' assessment, prediction, and forecasting workflows. 

\end{abstract}
%
%
%
\section{Introduction}
Massive data production is a critical aspect of experimental sciences.  It has not been different for geoscience. Examples of geoscientific data include any physical observable related to the energy industry, mining, monitoring hazardous areas (e.g. effects of salt mining in populated areas), etc. Nowadays, with the relative facility and lowering the cost to collect data, the data processing to exploit their value is a challenge. It requires expertise in data maintenance and processing, data analysis, and the design of experiments of target domains for which data will provide insight and knowledge.

This paper describes and discusses our vision to develop and reason about best practices and novel ways of curating  \cite{vargas2019demonstrating} data-centric geosciences knowledge (data, experiments, models, methods, conclusions, and interpretations). This knowledge is produced from applying statistical modelling, Machine Learning, and modern data analytics methods on geo-data collections. The problems address open methodological questions in model building, models' assessment, prediction, and forecasting workflows. 

This paper is organised as follows.
Section \ref{sec:relatedwork} discusses related work regarding existing disaggregated data centres approaches and data science workflow execution. 
Section  \ref{sec:approach} describes our vision and research challenges and opportunities of data centred smart geosciences. Section \ref{sec:usecases} describes examples of use cases addressed through data centred strategies using mathematical and Machine Learning or artificial intelligence algorithms. Section \ref{sec:conclusion} concludes the paper and discusses  future work.


\section{Related Work} \label{sec:relatedwork}
In France, portals like SISMER\footnote{\url{https://data.ifremer.fr/SISMER/Missions}} and Form@Ter\footnote{\url{https://www.poleterresolide.fr}} are initiatives willing to share data about target observation in geosciences and then share analytics experiments results. Data Terra\footnote{\url{https://www.data-terra.org}} is a research infrastructure dedicated to Earth System observation data.  In general, the objective of these platforms and portals is to facilitate access to satellite, airborne and in-situ data collected and managed by research laboratories or federative structures, by national infrastructures, the oceanographic fleet, aircraft, balloons, and by space missions (e.g., Data Terra). They manage, archive, and share TB of data. For example, Data Terra represented 50,000 TB in 2017 and is estimated to reach 100,000 TB by 2022. Beyond multi-source data, they also share products and services through a unified portal. Data is curated with metadata, included under accepted standards like the European standard INSPIRE. The challenge is to define common bases for all data producers and make the data sets interoperable so that their resources are consistent, shareable, exploitable, and, in a multidisciplinary approach, required to study the Earth. In this sense, the ODATIS Ocean Cluster offers several services for data producers similar to data labs for referencing, hosting, dissemination and interoperability. They also provide access to computing services for running experiments (models) that require important computing resources.

At the European level, actions adopting a data science perspective, for example, the project EPOS\footnote{\url{https://www.epos-ip.org}} and the Alan Turing Institute extend these initiatives to European partners willing to take full advantage of the possibilities provided by analytics and data science to run experiments and contribute to solving leading problems addressed by the discipline. Indeed, with the advent of digital technologies, libraries proposing analytics models have been run on mainframes and high-performance computing centres (HPC) to produce visualisation, modelling and simulation systems to accelerate interpretation and planning. 

Brazilian scientific agenda has widely installed and developed data centres like https://www.eveo.com.br/en/ and https://baxtel.com/data-center/brazil-brasil. These data centres aim to provide mainly large-scale computing resources to run experiments, for example, those regarding geosciences, particularly those key for the national economies in France and Brazil in oil and hydrocarbon exploitation, extraction of minerals, and its interaction with populated areas.

\section {Towards Smart Data Centred Geosciences}
\label{sec:approach}
Lately, geoscience researchers have been discovering the power of Machine Learning in solving problems in their field. Bergen et al. \cite{bergen2019machine}, for example, show that random forests were used on continuous acoustic emission in a laboratory shear experiment to model instantaneous friction and to predict time-to-failure \cite{rouet2017machine,kong2019machine}  surveyed the applications of Machine Learning in seismology and presented five research areas in in which Machine Learning classification, regression, clustering algorithms show promise: earthquake detection and phase picking, earthquake early warning
(EEW), ground-motion prediction, seismic tomography, and earthquake geodesy. In exploration geophysics, Machine Learning has been used in seismic data processing and reservoir characterization \cite{jia2017can,li2020seismic}. Clustering methods were used to identify key geophysical signatures and determine their relationship to rock types for geological mapping in the Brazilian Amazon \cite{carneiro2012semiautomated}. However, many researchers in the area are still not prepared to take advantage of data-driven approaches to their analyses at scale. In this context existing projects and actions are emerging to provide specialized portals and systems that can encourage the sharing of collected data (observations), experiments, and analytics results that should even promote reproducibility. 

In this context, we can see the emergence of multidisciplinary teams to collaborate in the search for computational solutions.
These teams are formed by experts in geology/geosciences, computer science, statistics and physics, among others.
The work of these teams usually relies on the use of mathematical/com\-pu\-ta\-tional tools to process large amounts of data.
Big data analytical techniques and Machine Learning has been used with success.

Many scientists and companies believe that they can generate fresh insight, reduce decision cycle times and steal a march on their competition by automating the search for patterns and relationships in their data. Therefore, geophysics and data science, including algorithms, mathematical models and computing, must converge for developing experiments for obtaining insight and foresight about the observations contained in data collections. Experiments represent best practices for addressing problems and questions on geophysics that must be treated as data and knowledge to be shared and reused by scientists and practitioners. 

Data collections issued in situ observations shared in pivot formats are vital for developing experiments that can lead to relevant governmental, economic, and social decision making. Information about how these data have been collected, used, curated, and maintained, including the conditions in which analyses are run and associated results and their use to lead to specific policies, should be managed and shared. 

  Merging data-centric techniques with modelling and simulation to answer questions in geoscience and make timely, clever, and disruptive decisions can lead to a new geoscience perspective that will benefit from data curation and analytics. In our vision, three important directions can be considered described in the following lines.

\paragraph{Collected data, models and knowledge integration.}
A wide variety of geophysical data (potential fields, electromagnetic data, seismic data, weather data, etc.) has been acquired with extensive wavelength ranges from surface sensor arrays, drilled wells, satellites and many other sources. These data sets are among the most significant science data sets in use, comparable in size and complexity only to those from astronomy and particle physics. Integrating access to data collections and their curated versions under a global knowledge graph can promote its maintenance, analysis, and experimentation. It can also show the knowledge of the discipline with its vocabulary, concepts, and relations in a synthetic manner. Inspired by existing public data labs like Kaggle or CoLab of Google, it can be essential to work to extend existing portals. These portals can be revisited towards specialized data science labs on geosciences. Through these labs, scientists and practitioners can share raw data, models, and experiments’ return of experience and run and reproduce other experiments with almost no requirement of interacting with specialized engineering support for accessing CPU and GPU clusters.

\paragraph{Curation, maintenance, exploration of data collections}
for bringing value to petabytes of data produced from in situ observations and also from experiments. Given that data act as a backbone in modelling phenomena for understanding their behaviour, it is critical to developing good collection and maintenance: which are available data collections? Are they complete? Which is their provenance? In which conditions were they collected? have they been processed? In which cases have they been used, and what are the associated results?

Data curation is a set of techniques to process (raw, distributed, heterogeneous) data to extract their value. It proposes methods to explore data collections using well-adapted data structures like graphs that can be explored and enriched while new data and analytics results are produced. Data curation means also keeping track of the type of experiments run on data, their results, and the conditions in which they were performed. 

Maintaining a catalogue of questions and experiments related to data can help provide a new vision of data and the scientific community's knowledge. This catalogue can extend existing meta-data and associated data collections information provided by actions like ODATIS and Data Terra.

\paragraph{Modelling and simulating experiments to answer questions in geoscience and make timely decisions.}
Both data sources and models come with recognized issues that existing methodologies have difficulties coping with but which novel data science-based approaches can address. For example, features for which exact physical models are unknown (e.g., subsurface geology, earthquakes) or models that are difficult to reconcile (e.g., seismic measurements vs social media alerts). This will imply:
\begin{itemize}
    \item 	Designing ad hoc experiment programming languages for enabling friendly, context-aware, and declarative construction of complex experiments in geosciences.
    
    \item Enabling the execution of experiments fusing different data collocations at different scales to maintain data, prepare experiments, and manage associated results.
    
    \item Programming experiments 
    \begin{itemize}
        \item 	Applying statistical methods to investigate and unveil new patterns in geophysical data, answering open problems, or leading to further research questions.
        \item Building predictive models to describe better or approximate geophysical phenomena, increasing the knowledge about our planet.
        \item Parallelizing algorithms for processing geophysical data, thus, allowing for the processing of very large data sets in reasonable times.
    \end{itemize}
\end{itemize}

\paragraph{Discussion}

From the Geophysics point of view, proposing best practices and ad-hoc strategies for developing data centred experiments to solve geosciences problems will impact different vital areas of the economy. For example, oil companies that ride this wave will significantly increase the current productivity of their knowledge workers, optimize business processes, and reduce operational costs in a way that is not possible through incremental change. Some companies now use algorithms to define optimal drilling locations, using automated or semi-automated systems that deliver results on much shorter cycle times than traditional methods.

From the Data Science/Data Processing perspective, this kind of multidisciplinary research can provide the ground to devise new data curation techniques, to propose a domain-specific query language, or to define new methods for processing heterogeneous data \cite{vargas2021towards}. In addition, statistical knowledge is essential for extracting information from the massive amount of data we will process. New methods and models will be crucial to model data and make conclusions in a timely fashion \cite{vargas2020enacting}.

\section {Use Cases}
\label{sec:usecases}
To illustrate the type of data analytics challenges introduced by geosciences problems, we describe in this section three examples. These use cases can be solved with different techniques and can call for data science strategies for specifying solutions and deploying them in target architectures.

\paragraph{Estimating the approximate earthquake epicentres.}
The understanding of earthquake occurrence in intraplate areas has been one of the most challenging tasks in Seismology \cite{fonseca2021intraplate}. 
Compared to border plate regions, interplate areas suffer less attenuation of seismic waves.
As a consequence,  a significant hazard may rise from moderate magnitude earthquakes.
Understanding the earthquake generating mechanisms depends on assessing the stress field in the intraplate areas.

\begin{figure}[ht!] \centering
\includegraphics[width=0.95\textwidth]{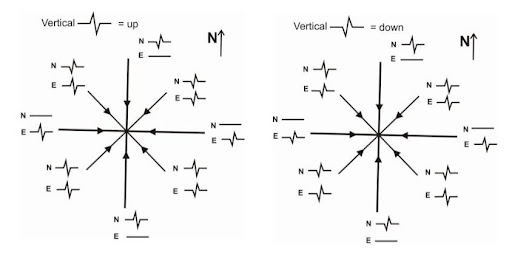}
\caption{Sensor movement.
\martin{Referencia?}
}\label{fig:sensor-movement}
\end{figure}

Seismic stations collect signals that can represent earthquakes produced in a specific area. The challenge is to determine whether signals represent earthquakes in such a case compute the epicentres.
For addressing the challenge, it is possible to use mathematical, Machine Learning or artificial intelligence methods \cite{leandro2020parallel}. The first task to address this question is to compute the earthquake epicentre's direction using the sensor's initial movement polarity when the waves P and S are discovered.
Then, compute the distance considering how the sensor moves from North-South (it should be the same as the East-West), as shown in Figure \ref{fig:sensor-movement}.

\begin{figure}[ht!] \centering
\includegraphics[width=0.95\textwidth]{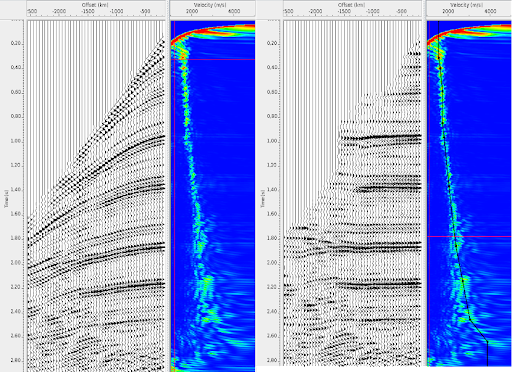}
\caption{NMO correction after velocity picking on the semblance.\martin{Referencia?}
}\label{fig:semblance}
\end{figure}

\paragraph{Estimation of stacking velocity using CDP semblance.}
Semblance analysis is a technique used in the study and refinement of seismic data. 
Along with other methods, this technique enables the improvement of the resolution of data, even in the presence of background noise. 
The data yielded by semblance analysis tends to be easier to interpret when discovering the underground structure of an area (see Figure \ref{fig:semblance}).

Estimating the stacking velocities is one of the essential steps in the CMP (Common Mid Point) seismic processing. This is because the better the estimation of the stacking velocities, the better the quality of the zero-offset section obtained. Currently, the most convenient velocity analysis method consists of manually picking the stacking velocities in the velocity spectrum, using the semblance as a coherence measure. The semblance gives us a measure of multichannel coherence.  It is necessary to define an analytics workflow with the following phases to perform this task: (i) transform the CDP or CMP gathered traces from the offset and time coordinates into the coherence semblances in coordinates of time and stacking velocities.  (ii)
Pick local maxima of these coherence semblances and assign zero offset time and corresponding stacking velocities. (iii) Correct the CDP or CMP gathers for normal moveout (NMO).

\paragraph{Denoising data from sensors.}
The Brazilian Seismographic Network (RSBR) operates since 2011. 
Station installation began in 2011 in southeast (SE) Brazil and finished in 2014 in the Amazon forest. 
The network integrates 84 stations (as of December 2017) operated by four institutions in different regions of Brazil. 
Seismic stations collect signals that can represent earthquakes produced in a specific area. This data usually contains noise produced by the context where the sensor is placed and by the technology of the sensor itself.
The challenge is to filter this data to make it ready to be analyzed.
This consortium is responsible for the Brazilian Seismic Bulletin~\cite{bianchi2018brazilian}. 

\section{Conclusions and Future Work}\label{sec:conclusion}
This paper proposes our vision about the multidisciplinary agenda for developing data centred solutions for geosciences problems. The amount of data collected through observing the Earth and its geophysical phenomena call for agile data and knowledge curation techniques to manage both data, experiments, and results.  The research agenda includes (i) integrating and describing data collected with different technology, (ii) estimating its quality, and preparing it to be used as input of different methods. Research on smart data centred geoscience also calls for curation tasks, including data tracking the way data is cleaned, the experiments that use it and the obtained results.  Exploration methods and systems must be associated with curated data and knowledge to facilitate an agile understanding of this content. Finally, execution environments providing computing resources necessary for setting and deploying experiments are vital for promoting multidisciplinary global experimental sciences. The research performed within the project ADAGEO \footnote{\url{https://adageo.github.io} funded by the IEA program of the French CNRS.} is willing to address these problems through a Brazilian and French collaborative community.

\bibliographystyle{splncs04}
\bibliography{biblio}
\end{document}